\documentclass[11pt,a4paper]{article}
\usepackage{amsmath,amssymb,mathrsfs}
\usepackage[dvips]{graphicx}
\usepackage{epsfig}

\def\un#1{\relax\ifmmode\@@underline#1\else
        $\@@underline{\hbox{#1}}$\relax\fi}

\let\du=\du                     

\def\a{\alpha}
\def\b{\beta}

\def\d{\delta}

\def\f{\phi}
\def\g{\gamma}
\def\h{\eta}

\def\m{\mu}
\def\n{\nu}

\def\p{\pi}

\def\r{\rho}

\def\z{\zeta}

\def\L{\Lambda}
\def\O{\Omega}

\def\Q{\Theta}

\def\ve{\varepsilon}

\def\cm{{\cal M}}

\def\bo{{\raise-.3ex\hbox{\large$\Box$}}}               
\def\pa{\partial}                                       
\def\TH{{\raise.2ex\hbox{$\displaystyle \bigodot$}\mskip-4.7mu \llap H \;}}
\def\face{{\raise.2ex\hbox{$\displaystyle \bigodot$}\mskip-2.2mu \llap {$\ddot
        \smile$}}}                                      

\def\VEV#1{\left\langle #1\right\rangle}        
\def\abs#1{\left| #1\right|}                    
\def\leftrightarrowfill{$\mathsurround=0pt \mathord\leftarrow \mkern-6mu
        \cleaders\hbox{$\mkern-2mu \mathord- \mkern-2mu$}\hfill
        \mkern-6mu \mathord\rightarrow$}
\def\dvec#1{\vbox{\ialign{##\crcr
        \leftrightarrowfill\crcr\noalign{\kern-1pt\nointerlineskip}
        $\hfil\displaystyle{#1}\hfil$\crcr}}}           
\def\dt#1{{\buildrel {\hbox{\LARGE .}} \over {#1}}}     

\def\sfrac#1#2{{\vphantom1\smash{\lower.5ex\hbox{\small$#1$}}\over
        \vphantom1\smash{\raise.4ex\hbox{\small$#2$}}}} 
\def\bfrac#1#2{{\vphantom1\smash{\lower.5ex\hbox{$#1$}}\over
        \vphantom1\smash{\raise.3ex\hbox{$#2$}}}}       
\def\afrac#1#2{{\vphantom1\smash{\lower.5ex\hbox{$#1$}}\over#2}}    

\def\[{\lfloor{\hskip 0.35pt}\!\!\!\lceil}
\def\]{\rfloor{\hskip 0.35pt}\!\!\!\rceil}

\def\du#1#2{_{#1}{}^{#2}}

\def\ha{{\fracmm12}}

\def\un{\underline}
\def\fracmm#1#2{{{#1}\over{#2}}}

\def\low#1{{\raise -3pt\hbox{${\hskip 0.75pt}\!_{#1}$}}}

\def\Dot#1{\buildrel{_{_{\hskip 0.01in}\bullet}}\over{#1}}
\def\dt#1{\Dot{#1}}

\newskip\humongous \humongous=0pt plus 1000pt minus 1000pt

\newif\ifdtup

\newcommand{\be}{\begin{equation}}
\newcommand{\ee}{\end{equation}}
\newcommand{\nbe}{\begin{equation*}}
\newcommand{\nee}{\end{equation*}}

\newcommand{\lb}{\label}

\def\lessim{\lower0.6ex\hbox{$\,$\vbox{\offinterlineskip\hbox{$<$}\vskip1pt\hbox{$\sim$}}$\,$}}
\def\grtsim{\lower0.6ex\hbox{$\,$\vbox{\offinterlineskip\hbox{$>$}\vskip1pt\hbox{$\sim$}}$\,$}}

\begin{document}

\begin{titlepage}

\begin{center}

July 2014 \hfill IPMU14-0022\\

\noindent
\vskip2.0cm
{\huge \bf 

Natural Inflation and Universal Hypermultiplet

}

\vglue.3in

{\large
Sergei V. Ketov~${}^{a,b,c}$ 
}

\vglue.1in

{\em
${}^a$~Department of Physics, Tokyo Metropolitan University \\
Minami-ohsawa 1-1, Hachioji-shi, Tokyo 192-0397, Japan \\
${}^b$~Kavli Institute for the Physics and Mathematics of the Universe (IPMU)
\\The University of Tokyo, Chiba 277-8568, Japan \\
${}^c$~Institute of Physics and Technology, Tomsk Polytechnic University\\
30 Lenin Ave., Tomsk 634050, Russian Federation \\
}

\vglue.1in
ketov@tmu.ac.jp

\end{center}

\vglue.3in

\begin{center}
{\Large\bf Abstract}
\end{center}
\vglue.1in

A novel framework is proposed for embedding the natural inflation into the type IIA superstrings compactified on a Calabi-Yau three-fold. Inflaton is identified with axion of the universal hypermultiplet (UH). The other UH scalars (including dilaton) are stabilized by the CY fluxes whose impact can be described by gauging of the abelian isometry associated with the axion, when the NS5-brane instanton contributions are suppressed. Then the stabilizing scalar potential is controlled by the integrable three-dimensional Toda equation, and leads to spontaneous N=2 SUSY breaking. The inflationary scalar potential of the UH axion is dynamically generated at a lower scale in the natural inflation via the non-perturbative quantum field effects such as gaugino condensation. The natural inflation has two scales that allow any values of the  CMB observables $(n_s,r)$.

\end{titlepage}


\section{Introduction}\label{sec:Intro}

The most economical, simple and viable inflationary models are the {\it single-field} quintessence field theories whose scalar potential is essentially controlled by a {\it single} parameter. Amongst the most popular models of that type are 
(i) the {\it Starobinsky} inflation \cite{star1,star2,mchi,star3,myrev}, the {\it Linde} inflation \cite{linde}, the {\it Higgs} inflation \cite{hi1,hi2,ks} and the {\it natural} inflation \cite{nat1,nat2,nat3}.

In the Starobinsky inflation, inflaton is scalaron (spin-0 part of metric). In the Higgs inflation, inflaton is identified with
a Higgs particle. In the natural inflation, inflaton is an axion. In the Linde inflation, the physical nature of inflaton is
unknown.

For instance, the Starobinsky inflation is based on the gravity action \cite{star1}
\be \lb{stara}
S[g] = \int \mathrm{d}^4x\sqrt{-g} \left[ -\ha R +\frac{1}{12M^2}R^2\right]~.
\ee
in terms of 4D spacetime metric $g_{\m\n}(x)$ with the scalar curvature $R$, where we have used the natural units
with the reduced Planck mass $M_{\rm Pl}=1$. Slow-roll inflation takes place in the high-curvature regime 
(with $M_{\rm Pl}>>H>> M$ and $|\dt{H}|<<H^2$), where the second term in Eq.~(\ref{stara}) dominates. Then the
Starobinsky inflationary solution (attractor!) takes the simple form 
\be \lb{ssol}
H\approx \frac{M^2}{6}(t_{\rm exit}-t)~,\qquad 0<t\leq t_{\rm exit}~~.
\ee
The inflationary model (\ref{stara}) has only one mass parameter $M$ that is fixed by the observational Cosmic 
Microwave Background (CMB) data as $M=(3.0 \times10^{-6})(\fracmm{50}{N_e})$ where $N_e$ is the e-foldings number. The predictions of the Starobinsky model for the spectral indices  $n_s\approx 1-2/N_e\approx 0.964$, $r\approx 12/N^2_e\approx0.004$ and low non-Gaussianity are in agreement with the WMAP and PLANCK data ($r<0.13$ and $r<0.11$, respectively, at 95\% CL) \cite{planck2}, but are in strong {\it disagreement} with the BICEP2 measurements ($r=0.2+0.07,-0.05$) \cite{bicep2}. 

The action (\ref{stara}) can be dualized by the Legendre-Weyl transform \cite{lwtr,myrev}  to the standard (quintessence)  action of the Einstein gravity coupled to a single physical scalar $\f$ having the scalar potential
\be \lb{starp}
V(\f) = \fracmm{3}{4} M^2\left( 1- e^{-\sqrt{\frac{2}{3}}\f }\right)^2~.
\ee

The coupling constant in front of the $R^2$-action is dimensionless. It results in the rigid scaling invariance of the Starobinsky inflation in the high curavture $R$ (or in the large field $\f \to +\infty$)  limit. This scaling invariance is not exact for finite values of $R$, and its violation is exactly measured by the slow-roll parameters, in full correspondence to the observed (nearly conformal) spectrum of the CMB perturbations. Thus the flatness of the inflaton scalar potential implies an (approximate) shift symmetry of the inflaton field.

Similar observations apply to the Higgs inflation \cite{hi1}, with the non-minimal coupling of the Higgs field to the scalar curvature, which also lead to the approximate (rigid) scale invariance during slow roll and, in fact, {\it the same} scalar potential (\ref{starp}) for inflation --- see e.g. Ref.~\cite{ks}.

Perhaps, the simplest inflationary model with a {\it quadratic} scalar potential, that was proposed by Linde \cite{linde}, predicts $r\approx 8/N_e=0.16\left(\fracmm{50}{N_e}\right)$ which is in good agreement with the BICEP2 data. A
consistency of the natural inflation with the PLANCK and BICEP2 data was discussed in Ref.~\cite{nat3}.

In our paper we leave aside the issue of apparent tension between the PLANCK and BICEP2 data, and concentrate
on a UV completion of the inflationary models. Indeed, the current status of all those models is phenomenological, 
they do not rely on a fundamental theory of gravity, and it is unknown whether it is possible at all. For example, the Starobinsky model  (\ref{stara}) has the finite UV-cutoff given by the Planck mass \cite{Hertzberg:2010}. A quantized $(R+R^2)$ gravity is still non-renormalizable, while its inflationary solution can be easily destabilized by adding the higher-order curvature terms $R^n$ with arbitrary coefficients to the action (\ref{stara}). The same remarks apply to the other inflationary models. It demands to complement the phenomenological (down-up) approach to inflation by the fundamental (up-down or {\it ab initio}) approach based on the superstring theory as a theory of 
quantum gravity, from the first principles. It would automatically provide the UV completion and control over quantum
corrections in any viable string-derived inflationary model and, perhaps, provide an ultimate resolution between all
inflationary models. In the past this type of research was always hindered by the absence of a nonperturbative formulation of string theory and a huge variety of possibilities for the choice of inflaton field.

There is, however, an exception given by the {\it Universal Hypermultiplet} (UH) in the {\it Calabi-Yau} (CY) compactified type IIA superstrings. The UH is known to be present in {\it any} CY compactification, and it has the {\it gravitational} origin. Moreover, the fully {\it non-perturbative} (exact) description of the UH is possible, in principle,
in string theory, and is partially known (see, e.g. Ref.~\cite{alexr} for a recent review).  

The closed type II strings give the UV completion of quantized gravity, while the "closed string
gravity" consist of the closed string zero modes including {\it metric, dilaton and $B$-field}, all being universally coupled to other fields. Their effective action (after integration of the string massive modes) gives rise to the (modified) Einstein gravity including the higher-order curvature terms. Those terms in the perturbative string effective action  can be computed from either string amplitudes of the massless modes or their equations of  motion given by the vanishing RG beta-functions of the {\it Non-Linear Sigma-mode}l (NLSM) describing string propagation in a background of the massless modes  --- see e.g.  Refs.~\cite{books,bookc,bookn} for a systematic discussion and explicit results. However, the coefficients in front of all Ricci- and scalar- curvature  dependent terms in the perturbative gravitational string effective action are {\it ambiguous}, because they are defined around the vacuum with the vanishing Ricci tensor. To resolve the ambiguity, one needs a non-perturbative setup for strings. It is usually unavailable, but there are some exceptions where the crucial role is played by {\it extended} local supersymmetry. 
Actually, the N=2 extended local supersymmetry in the critical dimension D=10 is required for consistency of closed (type II) strings, while their CY compactification  gives rise to N=2 local supersymmetry in 4D spacetime. The corresponding low-energy string effective action is given by a matter-coupled N=2 supergravity, while its moduli space $\cm$ is the direct product $\cm_V\otimes\cm_H$ of the moduli space $\cm_V$ of $h_{1,1}$  N=2 vector multiplets and the moduli space $\cm_H$ of $(1+h_{1,2})$ hypermultiplets, in terms of the CY Hodge numbers $h_{1,1}$ and $h_{1,2}$ (the UH is represented by $1$ in the $(1+h_{1,2})$ here).

Our motivation for this paper is to find the inflationary model that can be embedded (and thus UV-completed) in
string theory. We argue that the {\it natural} inflation can be embedded into the type IIA string theory compactified on
a CY three-fold, with inflaton-axion belonging to the universal hypermultiplet. The validity of this proposal requires
that the other (three) UH scalars, including dilaton, have to be stabilized.

First, let us recall that inflaton can be interpreted as the {\it pseudo-Nambu-Goldstone boson} (pNGb) $B$ associated with spontaneous breaking of the rigid scale invariance --- see e.g. Refs.~\cite{ks,nem}. It equally applies to the natural inflation too \cite{nat1}. When $f$ is a scale of spontaneous breaking of the scale invariance, and $\L$ is a scale of inflation, a typical  pNGb scalar potential takes the form \cite{nat1,nat2}
\be \lb{natp}
V(B) = \L^4\left[ 1 -\cos \left( \fracmm{B}{f}\right) \right]~.
\ee
In string theory, $f$ is of the order of the $M_{\rm Pl}$, whereas $\L$ originates in particle physics {\it dynamically},
via gaugino condensation \cite{gcond}. Our main proposal is to identify the axion $B$ of the natural inflation with the B-field of the UH in 4D.

Our paper is organized as follows. In Sec.~2 we briefly review the classical UH in the CY-compactified 4D, N=2 closed string theory. In the Sec.~3 we focus on the non-perturbative UH moduli space with the string loop  and
the D-instanton contributions included. In Sec.~4 we consider the UH scalar potential generated by the CY fluxes
via the gauging procedure, and emphasize the crucial role of the 3D Toda integrable system that controls the stabilizing scalar potential  of the UH scalars different from the B-axion (inflaton). It provides the simple ground for embedding the natural inflation into string theory.

\section{Classical UH moduli space}\label{sec:UHclassical}

The hypermultiplet moduli space $M_H$ of the CY-compactified 4D, type-IIA closed strings is known to be independent upon the CY complex structure but can receive non-trivial quantum corrections. The perturbative corrections are only possible at the 1-loop string level, being proportional to the CY Euler number  \cite{ant,ang}. The
non-perturbative (instanton)  corrections are due to the Euclidean D2-branes wrapped about the CY special  (supersymmertic) 3-cycles and due to the solitonic (NS-type) Euclidean 5-branes wrapped about the entire CY space. The 4D instantons due to the wrapped D2-branes are called {\it D-instantons} in the literature \cite{bbs,bb}.

By definition, a supersymmetric CY-cycle saturates the BPS bound, $J|_{C_3}={\rm Im}\O|_{C_3}=0$, where
$J$ is the CY K\"ahler form and $\O$ is the CY holomorphic 3-form, so that $C_3$ preserves half of the supersymmetry.  In the matematical literature \cite{doug} the supersymmetric  cycles are known as the special Lagrange sub-manifolds of the minimal volume in the given homology class,
$Vol(C_3)=\abs{\int_{C_3}\O}e^{-K}=minimal$ with the CY K\"ahler potential $K$.  

The semi-classical analysis of the wrapped D2-branes and 5-branes was given in Ref.~\cite{wi}, where it was found that the D-instanton corrections are of the form $\exp(-1/g_{ \rm string})$, whereas the wrapped 5-brane corrections have the form $\exp(-1/g^2_{ \rm string})$, in terms of the string coupling constant $g_{ \rm string}$. Therefore,
when $g_{ \rm string}$ is small, $g_{ \rm string}\ll 1$, the NS5-brane instanton corrections can be ignored.

For our purposes it is important to emphasize that all those corrections have the (super)gravitational origin  and respect local 4D, N=2 supesymmetry. It means that the metric of the UH moduli space is {\it quaternionic-K\"ahler}  \cite{bagw}. In other words, the kinetic terms of the UH in the type-IIA CY-compactified string theory are described by the NLSM having four real scalars and a  quaternionic-K\"ahler NLSM metric. 

The classical UH metric can be derived by compactifying the standard 10D type-IIA supergravity action down to 4D,
with the CY-Ansatz \cite{fsab}
\be \lb{compact}
ds^2_{10} = e^{-\f/2} ds^2_{\rm CY} + e^{3\f/2} g_{\m\n} dx^{\m}dx^{\n}
\ee
in terms of 4D dilaton field $\f(x)$. In the classical approximation one can replace a CY by (flat) torus with
$g_{ij}=\d_{ij}$ and $\O_{ijk}=\ve_{ijk}$ and keep only $SU(3)$ singlets ($i,j,k=1,2,3$): dilaton $\f$, axion $B$
arising from dualizing the 3-form field strength of the $B$-field in 4D, and a complex RR (Ramond-Ramond) scalar $C$ arising from the type IIA 3-form $A_3\propto C(x)\O$ in 10D. Then the 4D kinetic terms are given by \cite{fsab}
\be \lb{sab}
(\sqrt{-g})^{-1}L_{\rm cl.} = -\ha R - \ha(\pa_{\m}\f)^2 - \ha e^{2\f} \abs{\pa_{\m}C}^2 -  
\ha e^{4\f} \left( \pa_{\m}B + \frac{i}{2} \bar{C} \dvec{\pa}_{\m}C\right)^2~.
\ee
After the change of variables as $S=e^{-2\f}+2iB+\bar{C}C$ the NLSM metric in Eq.~(\ref{sab}) reads
\be  \lb{me1}
ds^2_{\rm cl.UH}= e^{-2K} \left( dSd\bar{S}-2C dS d\bar{C}-2\bar{C}d\bar{S}dC+2(S+\bar{S})dCd\bar{C}
\right)
\ee
with the K\"ahler potential 
\be \lb{ckp}
K =- \ln (S + \bar{S} - 2 C\bar{C} ) ~.
\ee
The further change of variables as
\be \lb{chvar}
z_1 = \frac{1-S}{1+S} \quad {\rm and} \quad z_2 = \frac{2C}{1+S} 
\ee
yields the most symmetric form of the metric as
\be \lb{me12}
ds^2_{\rm cl.UH}= \frac{ dz_1d\bar{z}_1 + dz_2d\bar{z}_2}{1-\abs{z_1}^2- \abs{z_2}^2 }
+\frac{ (\bar{z}_1dz_1 + \bar{z}_2dz_2)(z_1d\bar{z}_1 + z_2d\bar{z}_2)}{
(1-\abs{z_1}^2- \abs{z_2}^2){}^2 }~~,
\ee
that is known in the mathematical literature as the standard (Bergmann) metric in the open ball
$B^4:~\abs{z_1}^2+ \abs{z_1}^2<1$ in $\mathbb{C}^2$, with the (Fubuni-Study)  K\"ahler potential
\be \lb{fbkp}
K =- \ln (1- \abs{z_1}^2- \abs{z_1}^2 )~. 
\ee
The non-compact homogeneous space $B^4$ with the Bergmann metric is just the symmetric 
quaternionic coset space
\be \
 \cm_{\rm classical~UH}  = \frac{ SU(2,1)}{ SU(2) \times U(1)}~~.
\ee

In the string theory parametrization (\ref{me1}) the UH physical scalars are dilaton $\f$, axion $B$, and complex RR-field $C$.  The classical UH moduli space has three shift-like or PQ (Peccei-Quinn) symmetries \cite{fsab,bbs,bb}:
 \be \lb{pgs}
B\to B+\a~,\quad C\to C+ \g -i\b~,\quad S\to S +2(\g+i\b)C+\g^2+\b^2~,
\ee
with three real parameters $(\a,\b,\g)$, which generate a Heisenberg algebra.
 Those symmetries do not allow any non-trivial scalar potential for the axion $B$ and the RR field $C$, and, hence (by supersymmetry), for dilaton $\f$ also. Since (some) of those symmetries survive in the string (loop) perturbation theory, the scalar potential for the UH  in string theory can be generated only non-perturbatively (Sec.~4).

\section{Quantum UH moduli space}\label{sec:UHquantum}

In quantum 4D, N=2 closed string theory the non-perturbative UH moduli space is different from
the classical UH space (\ref{compact}),  as regards both its topology and its metric, because of the non-perturbative
d.o.f. in 4D due to the wrapped branes, and because some UH scalars get the non-vanishing VEVs in quantum theory that break the classical symmetries. In addition, the CY flux quantizaton implies quantized brane charges that can be identified with the Noether charges of the PQ symmetries (\ref{pgs}).  It is expected that the string duality
symmetry  described by the {\it discrete} group $SL(2,\mathbb{Z})$ always survives (though may be hidden on the
type IIA side).

Still, some of the classical continuous symmetries (\ref{pgs}) may survive in particular regions of the UH moduli space. For instance, as regards the D-instantons, the abelian isometry associated with shifts of the {\it axion} field survives, whereas even in a generic situation (when both D-instanton and 5-brane contributions included) a single 
isometry may also survive, as was demonstrated by an explicit example in Ref.~\cite{ang}. The residual isometry exist when one of the D-instanton or NS5-brane instanton charges (or a linear combination of them) vanishes. 
According to the Yano theorem \cite{besse}, the presence of any continuous isometry implies that the UH moduli space is non-compact and, hence, may be incomplete. The non-vanishing instanton charges break the symmetries
(\ref{pgs}) too. Those are the reasons against isometries in a {\it generic} region of the quantum UH moduli space.

The cosmological inflation can be associated with a special region of the quantum UH moduli space. We identify
that region by demanding the smallness of the string coupling, where the NS5-brane instantons are suppressed
(Sec.~2) and the axion isometry is preserved. Unlike the D-instantons, the NS5-brane instantons have
no analogues in quantum gauge theories.

The quantum gravity corrections are encoded in the quaternionic-K\"ahler structure of the quantum UH moduli space.
When assuming survival of a single isometry, the appropriate framework is given by a reformulation of the UH quaternionic-K\"ahler geometry as the {\it Einstein-Weyl} geometry with a {\it negative} scalar curvature, defined by the conditions \cite{besse}
\be \lb{ewc}
W^-_{abcd}=0~,\qquad R_{ab}=\frac{3}{2}\Lambda g_{ab}~,\qquad\Lambda=const. < 0~~,
\ee
where the $W_{abcd}^-$ is the anti-self-dual part of the Weyl tensor, and the $R_{ab}$ is the Ricci tensor of the 
UH moduli space metric $g_{ab}$, with $a,b=1,2,3,4$. Given the abelian isometry of the UH metric described by
a Killing vector $K$ obeying the equations
\be \lb{kill}
 K^{a;b} + K^{b;a} =0~, \quad K^2 = g_{ab}K^aK^b\geq 0~,
\ee
one can choose some adapted coordinates, in which all the metric components are
independent upon one coordinate $(t)$. Then the Przanowski-Tod theorem \cite{prz,tod} states
that any such metric with the Killing vector $\pa_t$ can be brought into the form 
 \be \lb{tode}
 ds^2_{\rm Tod}= \fracmm{1}{\r^2}\left\{ \fracmm{1}{P}(dt+\hat{\Q})^2 
+P\left[ e^u(d\m^2+d\n^2)+d\r^2\right]\right\}
\ee
in terms of the two potentials, $P$ and $u$, and the 1-form $\hat{\Q}$, 
in local coordinates $(t,\r,\m,\n)$. 

It follows from Eq.~(\ref{ewc}) that the potential $P(\r,\m,\n)$ is fixed by the 
second potential $u$ as \cite{tod} 
\be \lb{tod1}
 P= \fracmm{1}{\abs{\Lambda}} \left(1-\ha\r\pa_{\r}u\right)~~,
\ee
whereas the potential $u(\r,\m,\n)$ obeys the 3D {\it non-linear} equation 
\be \lb{toda}
-(\pa^2_{\m}+\pa^2_{\n})u+\pa^2_{\r}e^{-u}=0 \ee
that is known as the (integrable) $SU(\infty)$ or 3D continuous {\it Toda system}. Finally,
the 1-form $\hat{\Q}$ satisfies the {\it linear} differential  equation  \cite{tod} 
 \be \lb{tod2}
-d\wedge\hat{\Q} =(\pa_{\n}P) d\m\wedge d\r +(\pa_{\m}P) d\r\wedge d\n 
+\pa_{\r}(Pe^{-u}) d\n\wedge d\m~,
\ee
whose integrability condition is just given by Eq.~(\ref{toda}).
The classical UH metric (Sec.~2) in the parameterization (\ref{tode}) is obtained by taking
\be \lb{clat}
 P=\frac{3}{2\abs{\Lambda}}=const.>0~,\quad e^{-u}=\r~,\quad {\rm and}\quad d\wedge \hat{\Q}
= d\n\wedge d\m~~.
\ee
so that $u=2\f$. The string coupling is given by the dilaton VEV as $g_{\rm string}=\VEV{e^{\f}}$. 
The classical region  of the UH moduli space corresponds to the vanishing $g_{\rm string}$. 
The classical singularity appears at $\r\to 0$. 
 
The quantum UH moduli space was investigated in Refs.~\cite{ke1,ke2,ke3,ke4,bem,uhe2,uhe3}. 
In particular, as was found in Refs.~\cite{ke3,uhe2}, a partial summation of the D-instanton contributions is possible when  there is the extended $U(1)\times U(1)$ isometry. In this case the UH metric is governed by the Calderbank-Petersen potential $F(\r,\h)$ obeying the {\it linear} 2nd-order diferential equation \cite{cp}
\be \lb{calpe}
\r^2\left(\pa^2_{\r}+\pa^2_{\h}\right)F =
\fracmm{3}{4}F~~.
\ee
Its unique $SL(2,\mathbb{Z})$ modular invariant solution is given by the Eisenstein series $E_{3/2}$. The asymptotical expansion of the Eisenstein series reveals a sum of the classical contribution proportional to
$\r^{-1/2}$, the perturbative string 1-loop contribution proportional to $\z(3) \r^{3/2}$, and the infinite sum of the
D-instanton terms indeed \cite{ke3}. However, it seems to be problematic to have two isometries for the UH, as well
as for inflation.

The solutions to the Toda equation (\ref{toda}) describing the D-instanton contributions to the UH metric with an
abelian isometry were found in Ref.~\cite{uhe2}. The five-brane instanton corrections were studied in the past in Ref.~\cite{uhe3} and more recently in Ref~\cite{aban}.

 The quantum UH moduli space metric does not have a K\"ahler potential. However, the Weyl rescaling of the UH metric as $g_{ab}\to\r^2g_{ab}$ relates it to a K\"ahler metric with the vanishing scalar curvature \cite{gau}.

We expect that in the very strong string coupling (Landau-Ginzburg) region an approximate (rigid) N=2 superconformal symmetry appears. 

\section{CY fluxes and gauging the UH axion isometry}\label{sec:Gauging}

In the previous two Sections no scalar potential was generated for the UH scalars. As is well known in string theory, the moduli stabilization can be achieved  via adding non-trivial fluxes of the NS-NS and RR three-forms in CY 
\cite{dka}, while it amounts to {\it gauging} isometries of the UH moduli space in the effective 4D, N=2 supergravity \cite{ps}. As the abelian gauge field one can employ either gravi-photon of N=2 supergravity multiplet or a vector field of an N=2 matter (abelian) vector multiplet. As a result, the UH gets a non-trivial {\it scalar} potential whose critical points determine the vacua of the theory \cite{ps}. The gauging of {\it all} abelian isometries of the {\it classical} UH moduli space metric was investigated in Ref.~\cite{ita}, where it was found that it always leads to the run-away scalar potentials.

The scalar potential arising from the gauging procedure takes the form \cite{toy}
\be \lb{gpot}
 V = \frac{9}{2} g^{ab}\pa_a W\pa_bW - 6W^2\ee
in terms of the UH metric $g_{ab}$ and the superpotential $W$ defined by \cite{toy}
\be \lb{supot}
W^2 =\frac{1}{3}dK\wedge {}^*dK -\fracmm{1}{6}dK\wedge dK~,
\ee
where we have introduced  the Killing 1-form $K=k_adq^a$ of the gauged isometry and the Hodge star $(*)$ in any local coordinates $(q)$ on the UH moduli space. Unlike the N=1 local supersymmetry, the superpotential $W$ is not arbitrary but is fixed by the UH metric and the Killing vector.

In the parametrization of Eq.~(\ref{tode}) we have the Killing vector $K^a=(1,0,0,0)$ that yields the Killing 1-form
\be \lb{kform}
 K= \frac{1}{w^2P}(dt+\Q)~,\ee
whose square is given by
\be \lb{ksq}
K^2= g_{ab}K^aK^b=g_{tt}=\fracmm{1}{\r^2P}~~.
\ee

It is straightforward to compute the superpotential squared. We find ({\it cf.} Ref.~\cite{ke5})
\be \lb{spsq}
W^2= \fracmm{\L^2}{\r^2}+\fracmm{1}{12P}\left(3+2\abs{\L} P+
\fracmm{3}{2P}\r\pa_{\r}P\right)^2 
  +\fracmm{3\r^2e^{-u}}{4P^3}\left[ (\pa_{\m}P)^2+ (\pa_{\n}P)^2\right]
\ee

The first term in the scalar potential (\ref{gpot}) is always positive, whereas the second term is always negative, which is similar to the scalar potential in a generic matter-coupled N=1 supergravity \cite{crem}.  The Minkowski vacua are determined by the fixed points of the scalar potential, related to the poles of the function $P\r^2$ because of  Eq.~(\ref{ksq}). The existence of {\it meta-stable de Sitter} vacua was explicitly demonstrated in Refs.~\cite{bem,uhe2} in the presence of the D-instanton contributions.

The UH scalar potential (\ref{gpot}) in the parametrization (\ref{kform}) is non-negative provided that $P<4/3$. 
Getting an explicit example requires a non-separable solution to the 3D Toda equation (\ref{toda}) that is difficult to get 
 analytically (our numerical analysis will be given elsewhere). In the explicit examples of Ref.~\cite{ke5} some simplifying assumptions were made, as e.g. taking the hyper-K\"ahler limit or considering only D-instantons. It is just the last assumption that is enough for our purposes here, because it includes the quantum gravity corrections and  guarantees  survival of the axion isometry represented by the $t$-independence of the UH metric and the scalar potential above.

\section{Conclusion}

We proposed the inflationary scenario in the 4D quantum gravity given by the type IIA closed strings compactified on
a Calabi-Yau three-fold. Inflaton was identified with  the axion of the Universal Hypermultiplet.

The other (non-inflaton) scalars of the Universal Hypermultiplet (including dilaton) were stabilized by the CY fluxes
whose impact was calculated via the gauging procedure of the UH moduli space axion isometry. The latter survives
when the NS5-brane instantons are suppressed, i.e. at a small string coupling $g_{\rm string}\ll 1$.  A stabilization
of dilaton was one of the problems in the past, towards physical applications of the UH.

After the stabilization by CY fluxes/gauging, the N=2 local supersymmetry in 4D is spontaneously broken,  while
axion is still massless and has no scalar potential. However, at a lower scale the axion can get a scalar potential
due to some non-perturbative quantum field theory phenomena such as gaugino condensation. The slow-roll natural inflation can, therefore, take place with the the scalar potential (\ref{natp}) whose structure is essentially dictated by
the pNGb nature of the axion.  

It is worth noticing here that the scalar potential (\ref{natp}) of the natural inflation yields the scalar index $n_s$ and the tensor-to-scalar ratio $r$ of the CMB anisotropy as  \cite{nat1,nat2}
\be \lb{tilts}
n_s \approx  1- \fracmm{M^2_{\rm Pl}}{8\p f^2} \qquad {\rm and}\qquad \L \approx
2.2 \cdot 10^{16}~GeV \left( \fracmm{r}{0.002}\right)^{1/4}~.
\ee
Therefore, the CMB observables $(n_s,r)$ are directly related to the scales $(f,\L)$ of the natural inflation, respectively.

\section*{Acknowledgements}

The author thanks S. Alexandrov, E. Kiritsis, A. Sagnotti, A. A. Starobinsky and S. Vandoren for discussions. This work was supported by a Grant-in-Aid of the Japanese Society for Promotion of Science (JSPS) under No.~26400252, and the World Premier International Research Center Initiative (WPI Initiative), MEXT, Japan.
The author is grateful to the Laboratoire Charles Coulomb Universit\'e Montpellier 2 in France for hospitality
extended to him during the final stage of this work.


\begin{thebibliography}{99}
\bibitem{star1}   A.~A.~Starobinsky,  Phys.\ Lett.\ B {\bf 91} (1980) 99.
\bibitem{star2} A.~A. Starobinsky,  {\it Nonsingular model of the Universe with the quantum 
gravitational de Sitter stage and its observational consequences}, in the Proceedings of the 
2nd Intern. Seminar on “Quantum Theory of Gravity”, Moscow, 13–15 October 1981 (INR Press,
Moscow, 1982), p. 58; reprinted in "Quantum Gravity", eds. M.~A. Markov and P.~C. West 
(Plenum Publ. Co., New York, 1984), p. 103.
\bibitem{mchi}  V.~F. Mukhanov and G.~V. Chibisov, JETP Lett. {\bf 33} (1981) 532.
\bibitem{star3}   A.~A.~Starobinsky, Sov. Astron. Lett.  {\bf 9} (1983)  302.
\bibitem{myrev}  S.~V.~Ketov,  Int.\ J.\ Mod.\ Phys.\ A {\bf 28} (2013) 1330021; arXiv:1201.2239 [hep-th]. 
\bibitem{linde}  A. D. Linde, Phys.\ Lett.\ B {\bf 129} (1983) 177.
\bibitem{hi1} F. Bezrukov and M. Shaposhnikov, Phys. Lett. B 659, 703 (2008), arXiv:0710.3755 [hep-th].
\bibitem{hi2} F. Bezrukov, D. Gorbunov and M. Shaposhnikov, JCAP 0906 (2009) 029, arXiv:0812.3622 [hep-ph].\bibitem{ks} S.V.~Ketov and A.A.~Starobinsky, JCAP {\bf 08} (2012) 022, arXiv:1203.0805 [hep-th].
\bibitem{nat1} K. Freese, J. A. Frieman and A. V. Olinto, Phys. Rev. Lett. 65, 3233 (1990). 
\bibitem{nat2} F. C. Adams, J. R. Bond, K. Freese, J. A. Frieman and A. V. Olinto, Phys. Rev. D 47, 426 (1993) 
[hep-ph/9207245].
\bibitem{nat3}K. Freese and W. H. Kinney, {\it Natural infation: consistency with Cosmic Microwave Background
observations of Planck and BICEP2}, arXiv:1403.5277 [astro-ph.CO].
\bibitem{planck2} P.~A.~R.~Ade {\it et al.}  [Planck Collaboration],
 {\it Planck 2013 results. XXII. Constraints on inflation},  arXiv:1303.5082 [astro-ph.CO].
\bibitem{bicep2}  P.~A.~R.~Ade {\it et al.}  [BICEP2 Collaboration], {\it BICEP2 I: Detection of B-mode polarization at
degree angular scales},  arXiv:1403.3985 [astro-ph.CO].
\bibitem{lwtr} B.~Whitt, Phys. Lett. {\bf B145} (1984) 17.
 \bibitem{Hertzberg:2010} M.~P. Hertzberg, 
JHEP  {\bf 1011} (2010) 023; arXiv:1002.2995 [hep-ph].
\bibitem{alexr}  S. Alexandrov, Phys. Repts. {\bf 522} (2013) 1; arXiv:1111.2892 [hep-th].
\bibitem{books} S.~V. Ketov, {\it Introduction to Quantum Theory of Strings 
and Superstrings}, Nauka, 1990, Chapter 5 (in Russian).
\bibitem{bookc} S.~V. Ketov, {\it Conformal Field Theory},
World Scientific, 1995, Chapter VIII.
\bibitem{bookn} S.~V. Ketov, {\it Quantum Non-linear Sigma-models},
Springer-Verlag, 2000; Subsection 4.2.1.
\bibitem{nem} C. Cs\'aki, N. Kaloper, J. Serra and J. Terning, {\it Inflation from broken scale
invariance}, arXiv:1406.5192 [hep-th].
\bibitem{gcond} M. Dine, R. Rohm, N. Seiberg and E. Witten, Phys. Lett. {\bf B156} (1985) 55.
\bibitem{ant} I. Antoniadis, R. Minasian, S. Theisen and P. Vanhove,
Class. Quant. Grav. {\bf 20} (2003) 5079; arXiv:hep-th/0307268.
\bibitem{ang} L. Anguelova, M. Roc\v{e}k and S. Vandoren,
Phys. Rev.\ D {\bf 70} (2004) 066001; arXiv:hep-th/0402132.
\bibitem{bbs} K. Becker, M. Becker and A. Strominger,   Nucl. Phys.\ B {\bf 456} (1995) 130; arXiv:hep-th/9507158.
\bibitem{bb} K. Becker and M. Becker,   Nucl. Phys.\ B {\bf 551} (1999) 102; arXiv:hep-th/9901126.
\bibitem{doug} M. Douglas, {\it D-branes on Calabi-Yau manifolds}, arXiv:math.AG/0009209.  
\bibitem{wi} E. Witten,  Nucl. Phys.\ B {\bf 474} (1996) 343; arXiv:hep-th/9604030.
\bibitem{bagw} J. Bagger and E. Witten, Nucl. Phys.\ B {\bf 222} (1983) 1.
\bibitem{fsab} S. Ferrara and S. Sabharwal, Nucl. Phys.\ B {\bf 332} (1990) 317.
\bibitem{besse} A.~L. Besse, {\it Einstein Manifolds}, Springer-Verlag, 1987.
\bibitem{prz} M. Przanowski, J. Math. Phys. {\bf 32} (1991) 1004.
\bibitem{tod} K.~P. Tod, Lecture Notes in Pure and Appl. Math. {\bf 184} (1997) 307.
\bibitem{one} A. Strominger, Phys. Lett.\  {\bf 421} (1998) 139; arXiv:hep-th/9706195.
\bibitem{ke1} S.~V. Ketov, {\it D-instantons and universal hypermultiplet}, Los Angeles preprint 
CITUSC-01-046 (unpublished); arXiv:hep-th/0112012.
\bibitem{ke2} S.~V. Ketov,   Nucl.\ Phys.\ B {\bf 604} (2001) 256; arXiv:hep-th/0102099.
\bibitem{ke3} S.~V. Ketov,   Nucl.\ Phys.\ B {\bf 649} (2003) 365; arXiv:hep-th/0209003.
\bibitem{ke4} S.~V. Ketov,   Phys.\ Lett.\ B {\bf 558} (2003) 119; arXiv:hep-th/0302001.
\bibitem{bem} K. Behrndt and S. Mahapatra, J. High Energy Phys.  
 {\bf 01} (2004) 068; arXiv:hep-th/0312063.
\bibitem{uhe2} M. Davidse, F. Saueressig, U. Theis and S. Vandoren,
JHEP \  {\bf 0509} (2005) 065; arXiv:hep-th/0506097.
\bibitem{uhe3} S. Alexandrov, F. Saueressig and S. Vandoren,
JHEP \  {\bf 0609} (2006) 040; arXiv:hep-th/0606259.
\bibitem{aban} S. Alexandrov and S. Banerjee, {\it Fivebrane instantons in Calabi-Yau compactifications}, 
arXiv:1403.1265 [hep-th].


\bibitem{cp} D. M. J. Calderbank and H. Pedersen, {\it Self-dual Einstein 
metrics with torus symmetry}, math.DG/0105263.
\bibitem{gau} P. Gauduchon, {\it Surfaces k\"ahleriennes dont la courbure 
v\'erifie certaines conditions de positivit\'e}, in `G\'eom\'etrie 
riemannienne en dimension 4', S\'eminaire Besse 1978/1979, edited by 
L. B\'erard-Bergery et al., Textes Math. {\bf 3}, Paris, 1981.
\bibitem{dka} M.~R. Douglas and S. Kachru, Rev. Mod. Phys. {\bf 79} (2007) 733; arXiv:hep-th/0610102.
\bibitem{ps} J. Polchinski and A.  Strominger, Phys. Lett. {\bf 388B}  (1996) 736; arXiv:hep-th/9510227.
\bibitem{ita} A. Ceresole, G. Dall'Agata, R. Kallosh and A. Van Proeyen,
Phys. Rev.\ D {\bf 64} (2001) 104006; arXiv:hep-th/0104056.
\bibitem{ke5} S.~V. Ketov,   Nucl.\ Phys.\ B {\bf 656} (2003) 63; arXiv:hep-th/0212003.
\bibitem{toy} K. Behrndt and D. Dall'Agata, Nucl.Phys {\bf B627} (2002) 357; arXiv:hep-th/0112136.
\bibitem{crem} E. Cremmer, B. Julia, J. Scherk, S. Ferrara, L. Girardello and  P. van Nieuwenhuizen,
Nucl. Phys. {\bf B147} (1979) 105.

\end{thebibliography}
\end{document}